\documentclass[aps,prb,twocolumn,superscriptaddress,a4paper,floatfix,amsfonts,amsmath,amssymb,citeautoscript]{revtex4-2}

\usepackage{mathptmx}
\usepackage{float}
\usepackage[scaled=0.9]{helvet}
\usepackage[utf8]{inputenc}
\usepackage{graphicx}
\usepackage[pdftex,dvipsnames]{xcolor}
\usepackage[pdftex]{hyperref}
\hypersetup{
    pdftitle={},
    colorlinks,
    citecolor=blue
    }
\usepackage{upgreek}
\usepackage{bm}
\usepackage{textcomp}
\usepackage{xspace}
\usepackage{xargs}
\usepackage{verbatim}
\usepackage[normalem]{ulem}
\usepackage{ifthen}

\graphicspath{{./}{./figures/}}

\newcommand{\affiliationPDI}{\affiliation{Paul-Drude-Institut f\"ur Festk\"orperelektronik, Leibniz-Institut im Forschungsverbund Berlin e.V.,
Hausvogteiplatz 5-7, 10117 Berlin, Germany}}
\newcommand{\affiliationWIS}{\affiliation{Weizmann Institute of Science, Rehovot 76100, Israel}}

\newcommand{\fig}[2]{Fig.~\ref{#1}\ifthenelse{\equal{#2}{}}{}{(\lowercase{#2})}}
\newcommand{\eq}[1]{Eq.~\eqref{#1}}
\newcommand{\sect}[1]{Sec.~\ref{#1}}

\newcommand{\omegac}{\ensuremath{\omega_\text{c}}}
\newcommand{\rc}{\ensuremath{R_\text{c}}}
\newcommand{\lm}{\ensuremath{l_\text{m}}}
\newcommand{\vh}{\ensuremath{V_\text{H}}}
\newcommand{\ns}{\ensuremath{n_\text{s}}}
\newcommand{\rh}{\ensuremath{R_\text{H}}}
\newcommand{\rk}{\ensuremath{R_\text{K}}}
\newcommand{\mb}{\ensuremath{\mu B}}
\newcommand{\mbq}{\ensuremath{(\mu B)^2}}
\newcommand{\vd}{\ensuremath{\vec{\upsilon}_\text{d}}}
\newcommand{\vv}{\ensuremath{\vec{\upsilon}}}

\begin{document}

\title{Nature of current flow in the regime of the quantum Hall effect}

\author{Serkan Sirt}
\email{sirt@pdi-berlin.de}
\affiliationPDI
\author{Matthias Kamm}
\affiliationPDI
\author{Vladimir Y.\ Umansky}
\affiliationWIS
\author{Stefan Ludwig}
\email{ludwig@pdi-berlin.de}
\affiliationPDI

\begin{abstract}
The integer quantum Hall effect (QHE) belongs to the most fundamental phenomena of solid state physics and has an important application as resistance standard. It serves as a basis to understand the fractional, anomalous or spin QHEs, candidates for applications in quantum technology due to their topological properties. For optimizing all these applications it is essential to understand where the current flows inside the Hall bar, a question disputed for decades. We perform multiterminal current measurements on a Hall bar and compare the results with limiting models. We confirm, based on these experiments, that the current flow is chiral for the plateaus of quantized Hall resistance. Everywhere else, including the ranges between plateaus, our results comply with the Drude model, which predicts homogeneous current flow across a homogeneous Hall-bar.
\end{abstract}

\maketitle

\section{Introduction}\label{sec:introduction}

Magnetic field is an axial vector and, therefore, breaks time-reversal symmetry, which, however, can be restored by the simultaneous reversal of magnetic field and time, $\vec B \to -\vec B$ and $t\to-t$. This property gives rise to the chiral trajectory of a free charged particle in a homogeneous magnetic field, the mirror symmetry of the current in a two-terminal measurement, $I(\vec B)=I(-\vec B)$, or the point symmetry of the Hall voltage, $\vh(\vec B)=-\vh(-\vec B)$. In the regime of the quantum Hall effect (QHE), the situation is more complicated because of an inhomogeneous distribution of the electric field in case of a quantized Hall resistance. As a result, the current density distribution in the quantum Hall regime has been a matter of discussion \cite{Halperin1982,Buettiker1986,Buettiker1988,Fontein1991,Chklovskii1992,Geller1994,Zhitenev1994,Yahel1996,Mani1996,Yacoby1999,McCormick1999,Weitz2000,Dohi2007,Gerhardts2008,Siddiki2009,Siddiki2010,Weis2011,Kendirlik2013,Kendirlik2017,Gerhardts2020} since the discovery of the QHE \cite{vonKlitzing1980}.

In this article, we use multiterminal measurements to explore the implication of the axial symmetry of $\vec B$ for the current flow in the regime of the QHE. Multiterminal measurements are not capable of directly detecting the current density distribution, but based on a careful analysis they can nevertheless reveal the nature of current flow, e.g., if it is chiral or not. Before we discuss our experimental results, we introduce relevant theoretical aspects and models for comparison with our measurements.

In the classical equation of motion (EOM), $m\dot\vv=-e\vec E + \vec F_\text{L}$, the axial character of $\vec B$ is expressed in terms of the vector product of the Lorentz force $\vec F_\text{L}=-e\vv \times \vec B$ acting on an electron with charge $-e$ and effective mass $m$ at the momentary velocity \vv, exposed to a magnetic field and an electric field $\vec E$.

\begin{figure*}[tb]
\includegraphics[width=0.65\columnwidth]{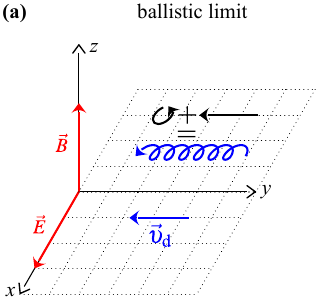}
\includegraphics[width=0.65\columnwidth]{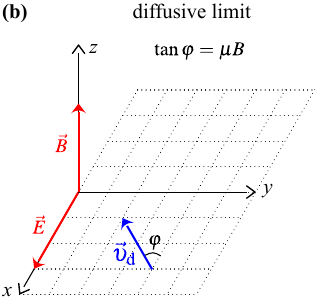}
\includegraphics[width=0.65\columnwidth]{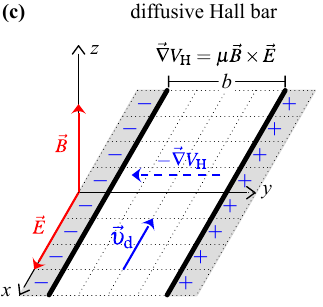}
\caption{Motion (expressed by drift velocity) of an electron confined to the $xy$-plane with in-plane electric and perpendicular magnetic field, $\vec E\perp\vec B$ (red arrows).
(a) Ballistic limit (free electron): The spiral indicates the cyclotron motion of an electron consisting of a uniform drift velocity $\vd$ and a (left-handed) circular motion with its radius being proportional to the initial velocity.
(b) Diffusive limit: the magnetic field reduces the drift velocity by the factor $1/\sqrt{(1+\mbq}$ compared to its absolute value for $B=0$ and tilts it by the Hall angle $\varphi$. The circular motion of the free electron is damped.
(c) Diffusive Hall bar with edges parallel to the applied $\vec E$: The Hall voltage compensates the Lorentz force, $-e\vec\nabla V_\text{H}=\vec F_\text{L}$, the drift velocity $\vd$ is parallel to the sample edges.
}
\label{fig:trajectory}
\end{figure*}

The solution provides the trajectory of a \emph{ballistic electron}. For $\vec E\not\perp\vec B$, it includes a constant acceleration $-e/(mB)(\vec E\cdot\vec B)$ along $\vec B$, where we denote the absolute value of the vector $\vec B$ as $B$. The more interesting part of the solution is a spiral-shaped trajectory confined to the plane perpendicular to $\vec B$ and composed of a uniform drift velocity $\vd=\vec E\times \vec B / B^2$ and a circular motion with radius $\rc=\upsilon_\perp/\omegac$, cf.\ \fig{fig:trajectory}{a}, where $\omegac=eB/m$ is the cyclotron frequency and  $\upsilon_\perp=||\vv\times\vec B||/B$ the (time independent) absolute value of the velocity component perpendicular to $\vec B$. If we define the chirality of this cyclotron motion in analogy to the handedness of a screw thread, it is left-handed (and would be right handed for a positive charge).

Next, we consider the \emph{diffusive regime}, which is dominated by momentum scattering. In its classical limit, its electron dynamics can be described by the Drude model with the electron mobility $\mu$ as central parameter. For $\mb\lesssim1$ (corresponding to a mean-free path $\lm\lesssim \rc$), momentum scattering counteracts the circular motion. Consequently, the Drude model omits the circular motion and replaces the velocity in the EOM (above) by the uniform drift velocity $\dot\vv\to\dot\vv_\text{d}$. After applying the relaxation time ansatz $\dot{\vv}_\text{d}=-\vd/\tau$, where $\tau=m\mu/e$ is the momentum scattering time, the EOM becomes a linear vector equation. Its solution,
\begin{equation}\label{Drude}
\vd = \frac{-\mu}{1+\mbq}\,\left[\vec E - \vec E \times \mu \vec B + \mu^2 (\vec E\cdot\vec B) \vec B\right]\,,
\end{equation}
can be simply determined by rearranging the EOM. For $\vec B\perp\vec E$ the term proportional to $\vec E\cdot\vec B$ vanishes and \vd\ is confined to a plane perpendicular to $\vec B$. Compared to $\vd=-\mu\vec E$ for $B=0$, a finite magnetic field reduces its absolute value according to $|\vd|=\mu E/\sqrt{1+\mbq}$. In respect to the direction of $\vec E$, the current density, $\vec j =-e \ns\vd$, is then bent by the Hall angle $\varphi=\sphericalangle(\vec j,\vec E)=\tan^{-1}(\mb)$, cf.\ \fig{fig:trajectory}{b}. Importantly, as long as both the carrier density, \ns,\ and the mobility, $\mu$, are homogeneous, $\vec j$ is homogeneous, too.

As argued above, the Drude model applies for $\mb\lesssim1$. Interestingly, for the opposite limit, namely $\mb\to\infty$ and $\rc\ll\lm$, the rapid circular motion of the cyclotron trajectory is not damped but averages out over reasonable length scales. At the same time, for $\mb\to\infty$, the friction forces in \eq{Drude} become negligible, such that the current density predicted by the Drude model becomes $\vec j=-e \ns\vd\to -e \ns \vec E\times \vec B / B^2$, which is independent of $\mu$ and identical to the solution for ballistic electrons.

In the remainder of the article, we restrict ourselves to a degenerate two-dimensional electron system (2DES), which defines the $xy$-plane, with an in-plane electric field $\vec E=(E_x,E_y,0)$ and a perpendicular homogeneous magnetic field $\vec B =(0,0,B)$. This is a common scenario realized in many mesoscopic magnetotransport measurements.

To discuss the \emph{classical limit of the Hall effect}, we include two sample edges at $y=0$ and $y=b$, such that the electrons are confined in the $y$-direction but move freely in the $x$-direction, cf.\ \fig{fig:trajectory}{c}. In the steady state, we expect $j_y=0$ everywhere even for $B\ne0$, because the edges reflect electrons. Applying $j_y=0$ to \eq{Drude} yields the classical Hall effect with $E_y=-\mb E_x$ and $j_x=E_y/\rh$, where we introduced the Hall resistance $\rh=-\frac{B}{\ns e}$. The transversal field $E_y$ is generated by charge accumulated at the sample edges, which can be measured in terms of the Hall voltage \vh, related to $E_y$ via $E_y=|\vec\nabla\vh|$. Because in steady state, the Lorentz force, acting on the carriers, is exactly canceled by $-eE_y$, the Drude model predicts that inside a Hall bar the electrons move as if only the longitudinal electric field $E_x$ was applied. Given a homogeneous sample (homogeneous distributions of \ns\ and $\mu$ away from the edges of the Hall bar), the current density $j_x$ is then predicted to be uniform and the current becomes $I=j_x b$, with $b$ being the width of the Hall bar.

Although the Drude model predicts a homogeneous current density for the classical Hall effect, the axial character of the magnetic field is actually manifest in the edge charges and the related Hall voltage, which break the time-reversal symmetry. Interestingly, other than an ohmic resistance, the Hall resistance does not add dissipation inside the Hall bar. (Remember, the related transversal field compensates the Lorentz force.) Instead, the Hall resistance is non-local. It causes additional dissipation where the current flow becomes parallel to $\vec\nabla\vh$, which typically happens near or inside the current carrying contacts. (For the quantum Hall regime, an enhanced dissipation near the contacts was experimentally confirmed \cite{Klass1991,Komiyama2006}). For a constant current, the overall dissipation is proportional to the two-terminal resistance, $R_\text{2T}$, of the Hall bar sample. At $B=0$, it is the sum of the resistances of the Hall bar and the leads (including the ohmic contacts), $R_\text{2T}(0)=R_0+R_\textrm{leads}$. At finite $B$, the resistance can be described as $R_\text{2T}(B)=\sqrt{R_0^2+\rh^2}+R_\textrm{leads}$.

As we increase $B$, at low enough temperatures the quantization of the density of states into discrete Landau-levels results in the \emph{quantum Hall effect}. Its main features are plateaus of quantized Hall resistance accompanied by a vanishing longitudinal resistance ($R_0\to0$). The latter indicates the absence of momentum scattering within the Hall bar for a quantized Hall resistance. The related Shubnikov-de-Haas (SdH) oscillations of the longitudinal resistance reveal in the regime of the QHE an oscillating $R_0(B)$ with growing maxima between the quantized plateaus as $B$ is increased, cf.\ \fig{fig:Hall_measurements}{} in Appendix \ref{app:c}.

To explain the quantized values of \rh, the Landauer-Büttiker picture (LBP) assumes the formation of chiral, one-dimensional (1D) edge channels, such that backward scattering is suppressed \cite{Buettiker1986,Buettiker1988}. The conductance of a spin-resolved 1D channel without momentum scattering is $e^2/h$ \cite{Landauer1981}, which concurs with the observed quantization of the Hall resistance, $\rh=\rk/\nu$, in fractions of the von-Klitzing constant $\rk=h/e^2$. The filling factor $\nu$ describes the number of edge channels and is equal to two-times the filling fraction of the (spin-degenerate) Landau levels in the 2D bulk. The LBP assumes that the edge channels form where the Landau levels, whose potential energies increase towards the depleted sample edges, intersect with the Fermi level  \cite{Buettiker1986,Buettiker1988}. To obtain finite ranges of quantized Hall resistances, additionally the bulk (between the edge channels) is assumed to be insulating due to localization in a disorder potential in a high magnetic field \cite{Buettiker1988}. The direction of the current flow in the 1D edge channels is predetermined by the gradient of the confinement potential. This yields current flow in opposite directions along opposite edges. Following the usual practice, we call this \emph{chiral current flow}. Depending on the direction of the magnetic field the chiral current flow is clockwise or counter-clockwise.

The LBP provides an intuitive explanation for the quantization of the Hall resistance and the accompanying vanishing longitudinal resistance of the QHE. While its merit is its simplicity, it neglects interactions and non-equilibrium effects, which gives it a phenomenological character with a limited range of validity. For example, the single particle assumption of the LBP leads to the prediction of a stepwise increase of the carrier density at the edges of the Hall bar because of the successive population of Landau levels \cite{Chklovskii1992}. It means, that neglecting the Coulomb interaction between carriers in the LBP causes the formation of highly charged stripes between the edge channels. This local polarization would give rise to an unrealistically large Coulomb energy. Another problem is the coexistence of occupied and unoccupied states at the chemical potential inside the current carrying edge channels according to the LBP. Even, if we take into account chiral current flow, an almost complete suppression of carrier scattering, as indicated by the high accuracy of the measurements of the von Klitzing constant \cite{Delahaye2003}, would then be unlikely under realistic conditions.

The screening theory \cite{Chklovskii1992,Chklovskii1993a,Fogler1994,Lier1994,Siddiki2003,Siddiki2004,Gerhardts2008} offers an alternative explanation of the QHE. By taking into account the direct Coulomb interaction between electrons, it overcomes the limitations of the LBP. It is based on the realization, that the Coulomb interaction of free carriers allows screening of the local electric field by means of rearrangement of carriers. This screening helps to avoid the highly charged stripes and thereby lower the free energy. However, the screening is hindered, if Landau-quantization in a strong magnetic field causes an energy gap. Early purely classical formulations included screening based on the Thomas-Fermi approximation \cite{Chklovskii1992,Chklovskii1993a}. By minimizing the free energy, they demonstrated that the Landau-quantization leads to a segmentation of the 2DES of a Hall bar in compressible regions and incompressible strips (ICSs). The term ICS denotes areas of the 2DES, where the chemical potential lies within a local energy gap between completely filled and empty Landau levels. For this reason, at low enough temperature, inside the ICSs carriers cannot scatter and screening of the electric field is absent. ICSs are accompanied by adjacent compressible regions, in which a partly filled Landau-level is exactly aligned with the chemical potential. Hence, inside compressible regions, carriers can scatter. Provided the vicinity of unscreened ICSs, the screening of the local electric field becomes perfect inside the compressible regions \cite{Chklovskii1992}. The early calculations already show, that both, width and position of the ICSs depend on the magnetic field as well as the local electron density distribution $\ns(x,y)$ at $B=0$. Later, for given boundary conditions, the number and geometry of the ICSs were predicted more accurately using a self-consistent numerical calculation of Poisson's equation within the Thomas-Fermi approximation and accounting for the quantum mechanical wavefunctions of the electrons in a mean field approach \cite{Siddiki2003,Siddiki2004,Gerhardts2008} (later confirmed in a fully self-consistent approach of the quantum-electrostatic problem \cite{Armagnat2020}). The calculations predict, that towards the higher magnetic field end of each plateau the ICSs merge into a single extended incompressible region (bulk ICS) centered in the Hall bar \cite{Siddiki2003,Siddiki2004,Gerhardts2008}. These calculations confirmed the electrostatic landscape measured in high-resolution scanning probe spectroscopy experiments as a function of the magnetic field \cite{Weitz2000,Ahlswede2001,Ahlswede2002}, which where indicated earlier by scanning electro-optical (Pockels) effect experiments in larger samples \cite{Fontein1991,Fontein1992}.

According to the screening theory the electric field vanishes inside the perfectly screened compressible regions while it is exclusively contained inside the ICSs. For suppressed scattering ($\mb\gg 1$) the drift velocity of the electrons is $\vd=\vec E\times \vec B / B^2$ (ballistic electrons). Hence, the current density, $\vec j =-e\ns\vd$, vanishes in the compressible regions, while all current flows inside the ICSs \cite{Chklovskii1993a,Guven2003,Gerhardts2008}. In case of a bulk ICS, current should then flow near the center of the Hall bar, while at the same time the Hall resistance should remain quantized. Recently, we  precisely confirmed this predicted bulk current by exploring current flow through an additional tiny ohmic contact in the center of a Hall bar \cite{Sirt2025a}. Both, the  electrostatics and the existence of bulk currents, predicted by the screening theory \cite{Gerhardts2008} and confirmed in measurements \cite{Fontein1991,Fontein1992,Weitz2000,Ahlswede2001,Ahlswede2002,Sirt2025a} are in direct contrast to the assumptions of the LBP. Therefore, it is important to find out, whether another assumption of the LBP, based on the axial symmetry of the magnetic field vector, remains correct, namely whether the current flow is chiral for the quantized plateaus of the Hall resistance and whether the chirality survives even in the case of bulk current.

The distribution of the local current density in the quantum Hall regime has been controversally discussed for some time, in particular, in articles going beyond the LBP \cite{Chklovskii1992,Chklovskii1993a,Mani1996,Yahel1996,Guven2003,Gerhardts2008,Weis2011,Gerhardts2020,Armagnat2020,Sirt2025a}. In the present article, we present experiments designed to explore the question of the chirality of the current flow by employing multiterminal current measurements. In detail, we probe the division of the current into various ohmic contacts of the Hall bar. Measuring currents into contacts, we cannot directly probe the local current density distribution. Nevertheless, it is possible to distinguish between chiral versus non-chiral transport as the current division into multiple contacts directly depends on it, see \sect{sec:4T-measurements} for related measurements.

To quantify the current distribution into various contacts, we use the properties of the macroscopic scattering matrix, which yields the division of the current into various contacts as follows: In the case of chiral current flow, only unidirectional transmission coefficients connecting adjacent contacts are non-zero. In comparison, non-chiral current flow requires additional non-zero transmission coefficients. Clearly, the two cases lead to different distributions of the current into the various contacts.

In \sect{sec:model} we introduce the two limiting models, the first one describing coherent and chiral current flow and the second one describing diffusive current (Drude model for multiple terminals). In \sect{sec:results}, we then present our three- and four-terminal current experiments and show that the measured current distribution into various contacts can be described for arbitrary magnetic fields by either one of two limiting models. Our results suggest, that the current flow is chiral for the quantized plateaus but unidirectional between the plateaus.

\section{models}\label{sec:model}

Our first model generically assumes coherent and chiral current flow. The condition of coherence is assumed for convenience. While this assumption is technically not necessary, the absence of scattering in an ICS would actually imply coherent current flow. The second model applies the classical Drude model to a multiterminal current measurement. It describes incoherent, diffusive and non-chiral current.

\subsection{Coherent and chiral model of \textit{n}-terminal device}\label{sec:coherent_model}

\begin{figure}[ht]
\includegraphics[width=.8\columnwidth]{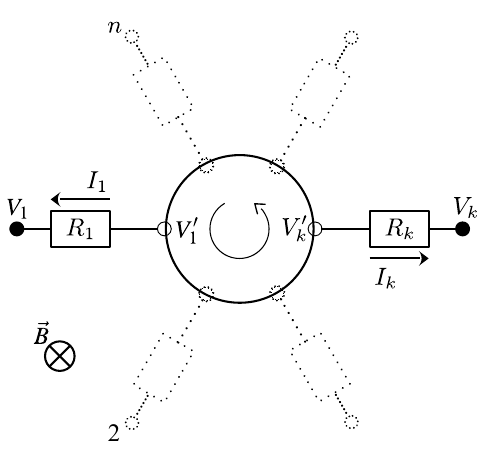}
\caption{Simplified circuit diagram of an $n$-terminal device. The inner circle indicates coherent and chiral current flow inside the 2DES. Each contact $k=1,...,n$ has an ohmic resistance $R_k$ and a voltage $V_k$ applied in respect to the electrical ground.  $V_k'$ are the voltages resulting at the intersections between the ohmic contacts and the coherent inner circle. The magnetic field vector pointing into the plane of the 2DES gives rise to a right handed chirality of a free charge which translates into a left handed chirality of the device, cf.\ circled arrow.}
\label{fig:nT-sketch}
\end{figure}

Our first model describes the limit of coherent and chiral current flow inside the 2DES of a sample equipped with $n$ ohmic contacts, cf.\ \fig{fig:nT-sketch}{}.
The inner circle in \fig{fig:nT-sketch}{} indicates the 2DES, the radial lines symbolize the $n$ terminals $k=1,2,...,n$ connected via ohmic contacts to the 2DES. External voltages $V_k$ can be applied to the contacts in respect to ground, while $V_k'=V_k+I_k R_k$ are the voltages present at the intersects of the 2DES and each contact. The current $I_k$ flows through the $k$th contact via the ohmic resistance $R_k$ to ground.

\begin{figure}[tb]
\includegraphics[width=1\columnwidth]{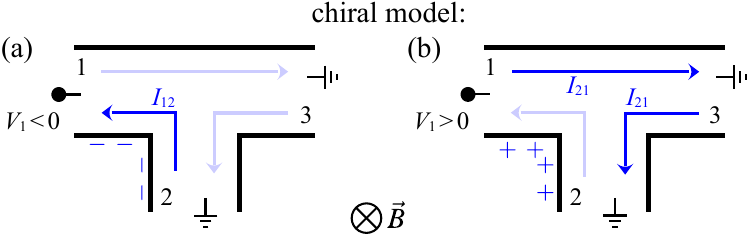}
\caption{Chiral current flow in a y-junction:
Simplifying sketches for a three-terminal junction for $\vec B$ pointing into the plane. In panel (a) a negative, in (b) a positive voltage $V_1$ is applied to contact 1, while the other contacts are grounded. Contact resistances are neglected, such that charge accumulates at the edge between contacts 1 and 2, only. Due to the direction of the applied magnetic field current flows clockwise (electrons counter clockwise). Bright blue arrows indicate the path way of the current generated by $V_1$. For $V_1<0$ it flows from contact 2 to 1; for $V_1>0$ it flows from 1 to 3 and then to 2. The faint arrows indicate a persistent chiral current that flows still for $V_1=V_2=V_3=0$.
}
\label{fig:chiral}
\end{figure}

To describe the current flow through the 2DES (inner part of the sample), we employ the Landauer-Büttiker formalism (not to be confused with the LBP) and write
\begin{equation}\label{LBF}
I_k=\sum_{j=1}^{n} G_{kj}V_j'=(V_k'-V_k)/R_k\,,
\end{equation}
where $G_{kj}$ are the conductance coefficients between mutual contacts. They can be formally derived from the scattering matrix of the problem. The left-hand side of \eq{LBF} is a general formulation of the transmission properties of the 2DES based on the Landauer-Büttiker formalism, while the last term of  \eq{LBF} applies Ohm's law to the macroscopic contacts. If the transport within the 2DES is chiral, only those $G_{k,j\ne k}$ connecting adjacent contacts can be non-zero. The direction of the chirality is linked to the direction of the magnetic field vector. In our case sketched in \fig{fig:nT-sketch}{}, we consider $\vec B$ pointing into the plane of the 2DES, such that $G_{kj}\ne0$ for $k=j+1$. Assuming coherent transport for the inner circle of \fig{fig:nT-sketch}{}, the scattering matrix is unitary, implying $\sum_k G_{kj} =\sum_j G_{kj}=0$, such that our coherent and chiral conductivity tensor takes the form
\begin{equation}\label{Gij}
G_{kj}=\nu G_0\left\{
\begin{matrix}
 1\,;   & k=j+1\\[1ex]
 -1\,;  & k=j\hfill\eject\\[1ex]
 0 \,;  & \text{else}\hfill\eject
\end{matrix}
\right.  \,,
\end{equation}
where  $G_0=e^2/h$ and $\nu$ is the filling factor of the (spin-resolved) Landau-levels. The assumption of coherence, which guarantees unitarity of the scattering matrix and the validity of \eq{Gij} is expected for the absence of scattering inside an ICS. For the quantized plateaus of the Hall resistance, $\nu$ is an integer. The \eq{Gij} might be considered as our formal definition of chiral current flow. Inserting \eq{Gij} into \eq{LBF} yields a solvable linear equation system providing us with the prediction of the current into the $k$th contact
\begin{equation}\label{currents_coherent}
I_k=\frac{1}{R_k}\left[\frac{1}{P_k}\left(\frac{S_n}{P_n-1}+S_k\right)-V_k \right]\,,
\end{equation}
where we used $V_k'=\frac{1}{P_k}\left(\frac{S_n}{P_n-1}+S_k\right)$ with
\begin{equation}
P_k=\left\{
     \begin{matrix}
      \prod_{j=1}^k \left(1+\frac{1}{\nu G_0 R_j}\right) \,; & k=1,2,...,n\\[1ex]
      1\,; & k=0\phantom{,2,...,n}
     \end{matrix}
     \right.\nonumber
\end{equation}
\begin{equation}
\text{and }S_k=\sum_{j=1}^k P_{j-1}\frac{V_j}{\nu G_0R_j}\,.\nonumber
\end{equation}
In Appendix \ref{app:a} we provide a derivation of \eq{currents_coherent}.
A simplifying special case of some relevance is that of $V_1\ne0$ but all other applied voltages $V_{i>1}=0$, which yields the current ratios
\begin{equation}
\frac{I_{k>1}}{I_1}=\frac{{P_n}/{P_k}}{1-{P_n}/{P_1}}\,\frac{1}{\nu G_0R_k}\,\nonumber
\end{equation}
irrespective of the sign of $V_1$. To visualize the nature of chiral current flow, in \fig{fig:chiral}{}
we sketch the chiral current path predicted by our model for just three terminals with $V_2=V_3=0$, while we neglect the contact resistances for simplicity. Arrows indicate the technical current direction, the electrons flow in the opposite direction. The clockwise chiral current flow is fixed by the direction of $\vec B$ pointing from the surface downwards through the 2DES plane. Contacts 2 and 3 are grounded, while $V_1<0$ in panel (a) and $V_1>0$ in (b). The chiral current predicted by \eq{Gij} includes a persistent contribution carried by states below the chemical potentials of all three leads, which flows even for $V_1=V_2=V_3=0$. Such a persistent current is expected for the plateaus of the QHE \cite{Chklovskii1993a,Guven2003}. The faint arrows indicate persistent current, while the bright blue arrows indicate the flow of the sum of the persistent current and the additional current imposed by applying $V_1\ne0$. For $V_1<0$, excess electrons move directly from contact 1 to contact 2 ($1\to2$) according to $V_1<V_2=V_3=0$, hence, the imposed current flows from $2\to1$. For $V_1>0$, the number of electrons flowing from $2\to3$ and that from $3\to1$ are equal, while the number of electrons flowing from $1\to2$ is smaller. Therefore, the imposed current flows from $1\to3\to2$.

In summary, the path of the chiral current is fixed by the direction of the magnetic field. As a result, if we neglect contact resistances, all current generated by applying a voltage to contact 1, always flows through contact 2 independent of the sign of $V_1$ or the number of additional grounded contacts. Upon adding contact resistances, \eq{currents_coherent} predicts corrections, such as some current flowing through additional contacts.

\subsection{Drude model of an \textit{n}-terminal device}\label{sec:Drude-nterminal}

The Drude model provides a solution of \eq{LBF}, which describes the classical limit of diffusive and incoherent currents inside the 2DES.
Using $\vec j = -e\ns\vd$ we can rewrite \eq{Drude} in terms of the magnetic-field-dependent conductivity tensor as
\begin{equation}
\begin{pmatrix} j_x \\ j_y \end{pmatrix} = \frac{\sigma_0}{1+\mbq}
\begin{pmatrix} 1 & -\mb \\ \mb & 1  \end{pmatrix}\,
\begin{pmatrix} E_x \\ E_y \end{pmatrix}\,,
\end{equation}
where the specific conductivity (defined at $B=0$) is $\sigma_0=e\ns\mu$. For a 2DES laterally confined in the $y$-direction, in the steady state we expect $j_y=0$, which implies $E_y=-\mb E_x$ and, consequently, $j_x=\sigma_0 E_x =\rh^{-1} E_y$.

The structure of the conductivity tensor allows us to partition the non-vanishing current density $j_x$ into two components with distinct physical meanings, namely
\begin{eqnarray}\label{jx}
&j_x \equiv j_{xx}+j_{xy} \quad\text{with}\\
&j_{xx}=\frac{\sigma_0}{1+\mbq}\,E_x\quad\text{and}\quad
 j_{xy}=\frac{\mbq}{1+\mbq}\,\frac{1}{\rh}\,E_y\,.\nonumber
\end{eqnarray}
The longitudinal component $j_{xx}=\sigma_{xx}E_x$ is the ohmic term, proportional to the specific conductivity and driven by the longitudinal electric field $E_x$. Caused by the additional Hall resistance, at finite $B$ the ohmic current density is reduced by the factor $\left[1+\mbq\right]^{-1}$ and vanishes for $\mb\gg1$. The transversal component $j_{xy}=\sigma_{xy}E_y$ is proportional to the perpendicular electric field. For  $\mb\gg1$, with $j_{xy}\to E_y/\rh$ it becomes independent of the ohmic resistance. This implies, that for $\mb\gg1$ momentum scattering is absent. At the same time, the transversal $j_{xy}$ follows the charged edge of the Hall bar. This way, it is bent into a current carrying contact.

\begin{figure}[tb]
\includegraphics[width=1\columnwidth]{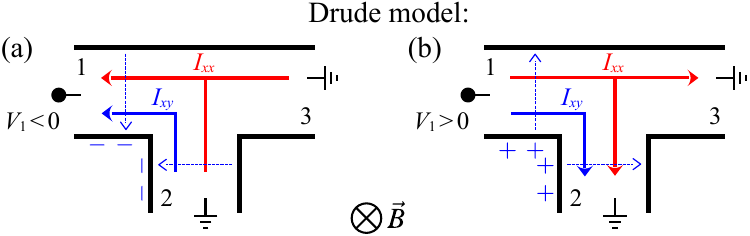}
\caption{Diffusive current flow in a y-junction (Drude model):
Simplifying sketches showing the partitioning of the current contributions $I_{xx}$ versus $I_{xy}$ in a three-terminal junction for $\vec B$ pointing into the plane. In panel (a) a negative, in (b) a positive voltage is applied to contact 1, while the other contacts are grounded. Contact resistances are neglected, such that charge accumulates at the edge between contacts 1 and 2, only. $I_{xx}$ is devided according to Kirchhoff's current law. $I_{xy}$ is guided by the charged edge and flows perpendicular to the Hall field $E_y$ (indicated by dashed arrows). $I_{xy}$ flows from contact 2 to 1 (opposite to electron flow).
}
\label{fig:Drude}
\end{figure}

In our experiments, we measure the current, which corresponds to the integral  across the Hall bar of the current densities in \eq{jx}. To illustrate the contributions of $I_{xx}$ and $I_{xy}$, in \fig{fig:Drude}{}
we sketch the current division according to the Drude model in a three-terminal junction for the identical conditions as in \fig{fig:chiral}{}, where we sketched current path ways within the chiral model, ($V_1\ne0$, $V_2=V_3=0$, $R_1=R_2=R_3=0$). $I_{xx}$ branches according to Kirchhoff's current law, while the non-ohmic $I_{xy}$ is bent by the edge charges (respective magnetic field) into the next available (non-floating) contact in current direction. The classical Drude model predicts unidirectional current flow, which is in no way chiral. Nevertheless, the current component $I_{xy}$ is bent such that it flows between contacts 1 and 2, as does the imposed current predicted by our chiral model in \sect{sec:coherent_model}.

Applying the Drude model to our general $n$-terminal sample allows us to predict the currents $I_k$ flowing into each contact in the diffusive regime. If we assume a homogeneous 2DES in our Hall bar and neglect possible asymmetries in the contact configuration, we may incorporate the resistance $R_0(B)$ of the 2DES into the $n$ contact resistances, and find
\begin{eqnarray}\label{currents_Drude_nterminal}
I_{k}&=&I_{xx}^{(k)} + I_{xy}^{(k)}\quad\text{with}\nonumber\\[1ex]
I_{xx}^{(k)}&=& \frac{1}{1+\mbq}\,\frac{1}{R_k}\left[\frac{\sum_{j=1}^{n}V_j/R_j}{\sum_{j=1}^{n}1/R_j}-V_k\right]  \\[1ex]
I_{xy}^{(k)}&=& \frac{\mbq}{1+\mbq}\,\frac{1}{R_k}\left[\frac{1}{P_k}\left(\frac{S_n}{P_n-1}+S_k\right)-V_k \right]\,  \nonumber
\end{eqnarray}
for the current flowing into the $k$th contact. In Appendix \sect{app:b} we provide a detailed explanation of this result and discuss the relevance of $R_0(B)$ for the case of an asymmetric configuration of contacts.

The original scope of the classical Drude model was a description of diffusive transport at $\mb\lesssim1$. However, in the limit $\mb\to\infty$, the first term of \eq{currents_Drude_nterminal} vanishes, $I_{xx}\to0$, and the prediction of the Drude model becomes identical to \eq{currents_coherent}, see Appendix \ref{app:b} for a discussion. Therefore, we can interpret \eq{currents_Drude_nterminal} as follows: for the limit $\mb\lesssim1$, which occurs for non-integer $\nu$ between the plateaus of the QHE, \eq{currents_Drude_nterminal} describes diffusive transport based on the Drude model. For the opposite limit $\mb\gg1$, which occurs for integer $\nu$ on the plateaus, \eq{currents_Drude_nterminal} precisely describes transport according to our chiral model. For intermediate \mb, \eq{currents_Drude_nterminal} smoothly interpolates between the limits. This is useful, because $\mu(B)$ strongly oscillates due to the SdH oscillations.

\section{experimental results and discussion}\label{sec:results}

Multiterminal current measurements yield the distribution of current into various contacts. In this section, we discuss the results of such experiments and compare them with theory predictions in a wide magnetic field range corresponding to filling factors between $1\lesssim\nu<\infty$.

\subsection{Sample and measurement setup}\label{sec:setup}

\begin{figure}[tb]
\includegraphics[width=.8\columnwidth]{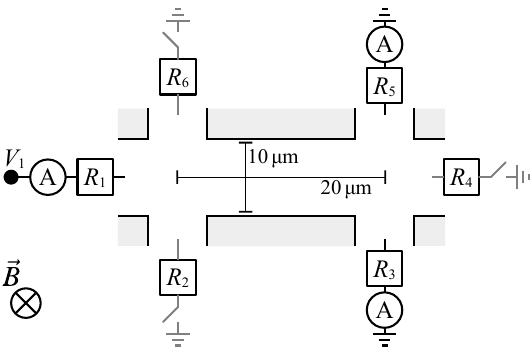}
\caption{Sketch of a Hall bar with six contacts, characterized by their ohmic resistances, $R_k$. We apply a voltage $V_1$ to contact 1 and measure currents into contacts 1, 3, and 5. Contacts 2, 4, and 6 can be left floating or connected to electrical ground. Current amplifiers at contacts 3 and 5 result in additional voltages of $\simeq20\mu$V applied to contacts 3 and 5 (not shown, but taken into account for generating model curves). $\vec B$ points into the plane of the 2DES.}
\label{fig:Hallbar}
\end{figure}

We use a Hall bar as sketched in \fig{fig:Hallbar}{}.
It is crafted by wet-etching from a GaAs\,/\,(Al,Ga)As heterostructure, which contains a two-dimensional electron system (2DES) 130\,nm beneath the surface. We performed the experiments at a temperature of $T\simeq300\,$mK in  a He-3 evaporation cryostat. At this temperature and zero magnetic field, the electron density and mobility of the 2DES, determined from Hall measurements, are $ n_\textrm{s} \simeq 1.2\times 10^{11} \textrm{ cm}^{-2} $ and $ \mu \simeq 3.95\times 10^6 \textrm{cm}^2/\textrm{Vs}$, respectively (yielding a mean-free-path of $\lambda_\text{m}\simeq23\,\upmu$m).

\begin{figure}[tb]
\includegraphics[width=1\columnwidth]{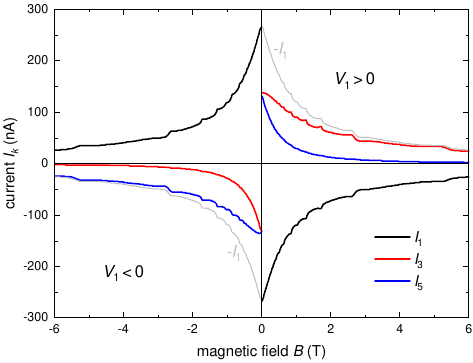}
\caption{Currents $I_k(B)$ measurement of a symmetric y-junction: Contacts 2, 4 and 6 in Fig.\ \ref{fig:Hallbar} are floating, $V_1=+1\,$mV for $B>0$ and $V_1=-1\,$mV for $B<0$, contacts 3 and 5 are connected to the electrical ground. Measured currents are $I_1$ (black line), $I_3$ (red) and $I_5$ (blue), gray lines correspond to $-I_1$. We find $I_3>I_5$ for $B>0$ and $I_5>I_3$ for $B<0$, where $I_1+I_3+I_5=0$. }
\label{fig:overview}
\end{figure}

For multi-terminal current measurements, we apply the voltage of $V_1=\pm1\,$mV to the source contact 1 (using a Keithley 2450 source meter). We measure the current $I_1(B)$ flowing from the Hall bar into contact 1, using the source meter, and the currents $I_3(B)$ and $I_5(B)$ flowing into contacts 3 or 5, respectively, using Basel Precision Instr.\ current amplifiers. Contacts 3 and 5 are connected to the electrical ground, while the remaining contacts 2, 4, and 6 may be either electrically floating or connected to ground depending on the experiment.
For a first impression, in \fig{fig:overview}{}
we present the measured currents $I_1(B)$, $I_3(B)$ and $I_5(B)$ for two three-terminal measurements (contacts 2, 4 and 6 are floating, such that contacts 3 and 5 correspond to contacts 2 and 3 in the sketches in \fig{fig:chiral}{} and \fig{fig:Drude}{}). We define $B>0$ for $\vec B$ pointing downwards (from the surface through the 2DES) and $B<0$ for $\vec B$ pointing upwards. We applied the voltages $V_1=+1\,$mV and swept $B>0$ versus $V_1=-1\,$mV for $B<0$. In both cases, for $B=0$, the currents $I_3\simeq I_5$, because the contact resistances are almost equal. (Of course, the sign of the currents depends on the sign of the voltage.)  As expected, $-I_1=I_3+I_5$ always  $|I_1|$ decreases as $|B|$ is increased because of $\rh\propto |B|$. Importantly, for $B>0$, we observe $|I_3|>|I_5|$, while for $B<0$, we find $|I_5|>|I_3|$. This behavior is independent of the sign of $V_1$ (not shown here). These data directly visualize our prediction that the current distribution in the various grounded contacts is governed by the direction of $\vec B$, only, cf.\ \eq{currents_Drude_nterminal} or the sketches in \fig{fig:chiral}{} and \fig{fig:Drude}{}.

The magnetic field dependences $|I_1(B)|$ are not exactly mirror symmetric in respect to $B=0$, as would be the case for a two-terminal measurement. The deviations point to slight differences between the contact resistances and are also influenced by the two current amplifiers connected to contacts 3 and 5, which have small offset voltages in the range of $V_{3,5}\sim10\,\mu$V varying between amplifiers. To determine the contact resistances and offset voltages, we performed a series of two-terminal measurements at $B=0$. In addition, we fine-tuned the values of these circuit parameters by comparing magnetic field dependent measurements with our models. These circuit parameters remained stable for the duration of all experiments discussed below. We list their values as we used them in all our model curves in Table \ref{table:parameters}.
\begin{table}[htb]
 \centering
 \begin{tabular}{|c|c|c|c|}
 \hline
 contact $k$    & resistance $R_k$ ($\Omega$)  &  voltage $V_k$ (mV) &  geometry $a_k$    \\\hline
 1              &  2420             &   1.000     & 1         \\\hline
 2              &  2420             &   ---       & 0.5       \\\hline
 3              &  2400             &  -0.021     & 1         \\\hline
 4              &  2350             &   ---       & 1         \\\hline
 5              &  2420             &  -0.018     & 1         \\\hline
 6              &  2400             &   ---       & 0.5       \\\hline
 \end{tabular}
\caption{Parameters of the circuit sketched in Fig.\ \ref{fig:Hallbar}. Contact resistances $R_k$ include ohmic contacts, cables and RC-filters. Voltage $V_{1}=1$\,mV is applied to contact 1, $V_3$ and $V_5$ are offset voltages of the current amplifiers connected to the respective contacts. Contacts 2, 4 and 6 are either floating or directly connected to ground. Hence, $V_2$, $V_4$ and $V_6$ are zero (variable), if the respective contact is grounded (floated). To account for the contribution of the Hall bar itself to the resistance, we defined a geometry parameter and use $R_k\to R_k+a_k/[\ns e \mu(B)]$ in the model calculations of the ohmic contributions; $a_2$ and $a_6$ are smaller, as they are closest to contact 1, such accounting for a shorter current path that comes into effect when the respective contacts are grounded. The correction is negligible for $B=0$.
}
\label{table:parameters}
\end{table}
This preparation allows us to perform quantitative comparisons between measurements and models. For the reminder of the article we discuss experiments with $V_1=1\,$mV and $B>0$. For the chiral model, electrons then move from contact to contact along the counterclockwise pathway $1\to2\to3\to4\to5\to6\to1$. Since the external voltage that we apply to contact 1 exceeds all other voltages by far (and all contact resistances are similar), for a sizable magnetic field $B>0$, according to the prediction of both models and our first experiment shown in \fig{fig:overview}{}, the largest current will flow between contacts 1 and the next grounded contact following in counterclockwise direction.

\subsection{Three-terminal measurements}\label{sec:3T_measurementss}

\begin{figure*}[tb]
\includegraphics[height=0.85\columnwidth]{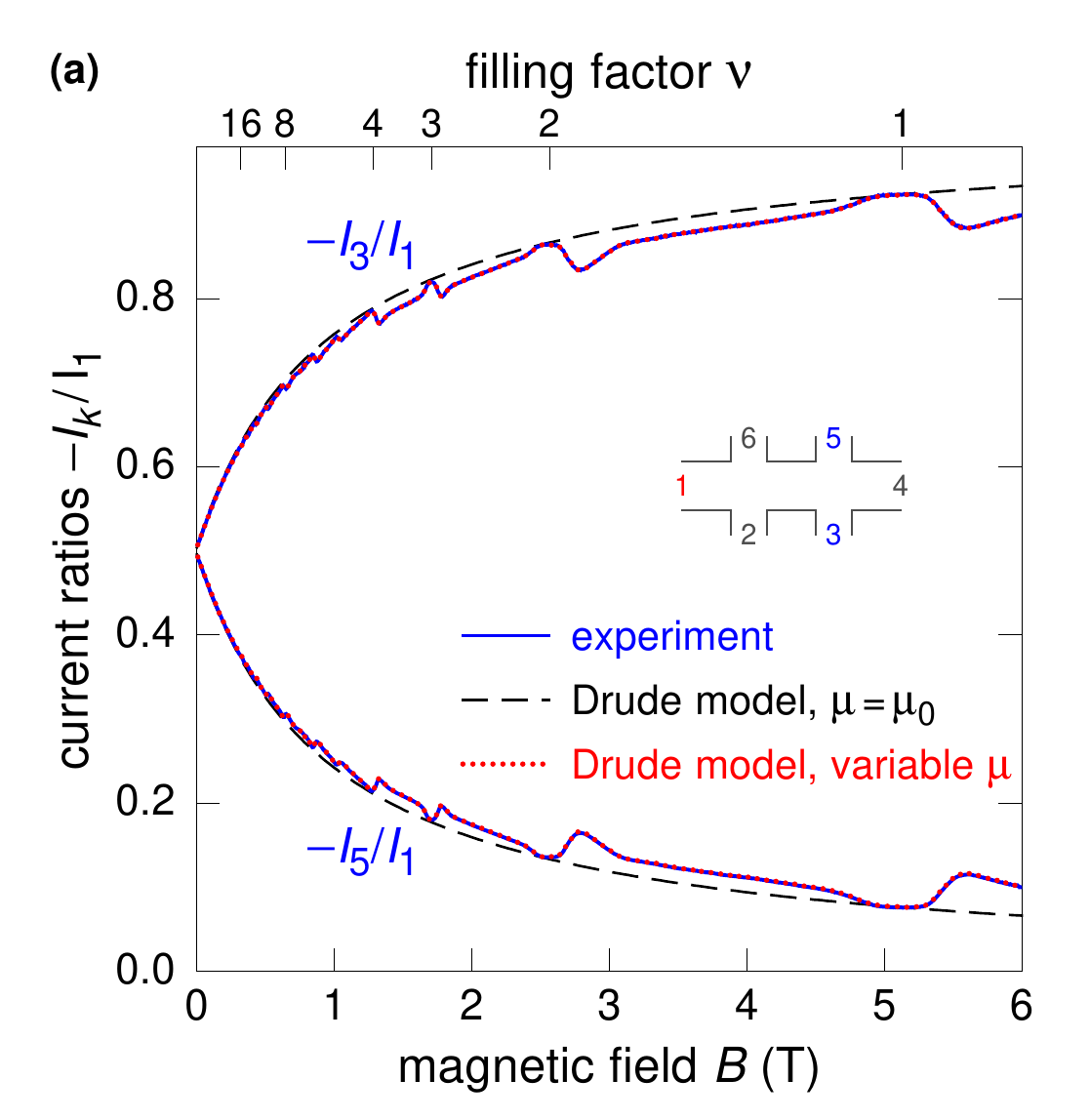}
\hspace{20mm}
\includegraphics[height=0.85\columnwidth]{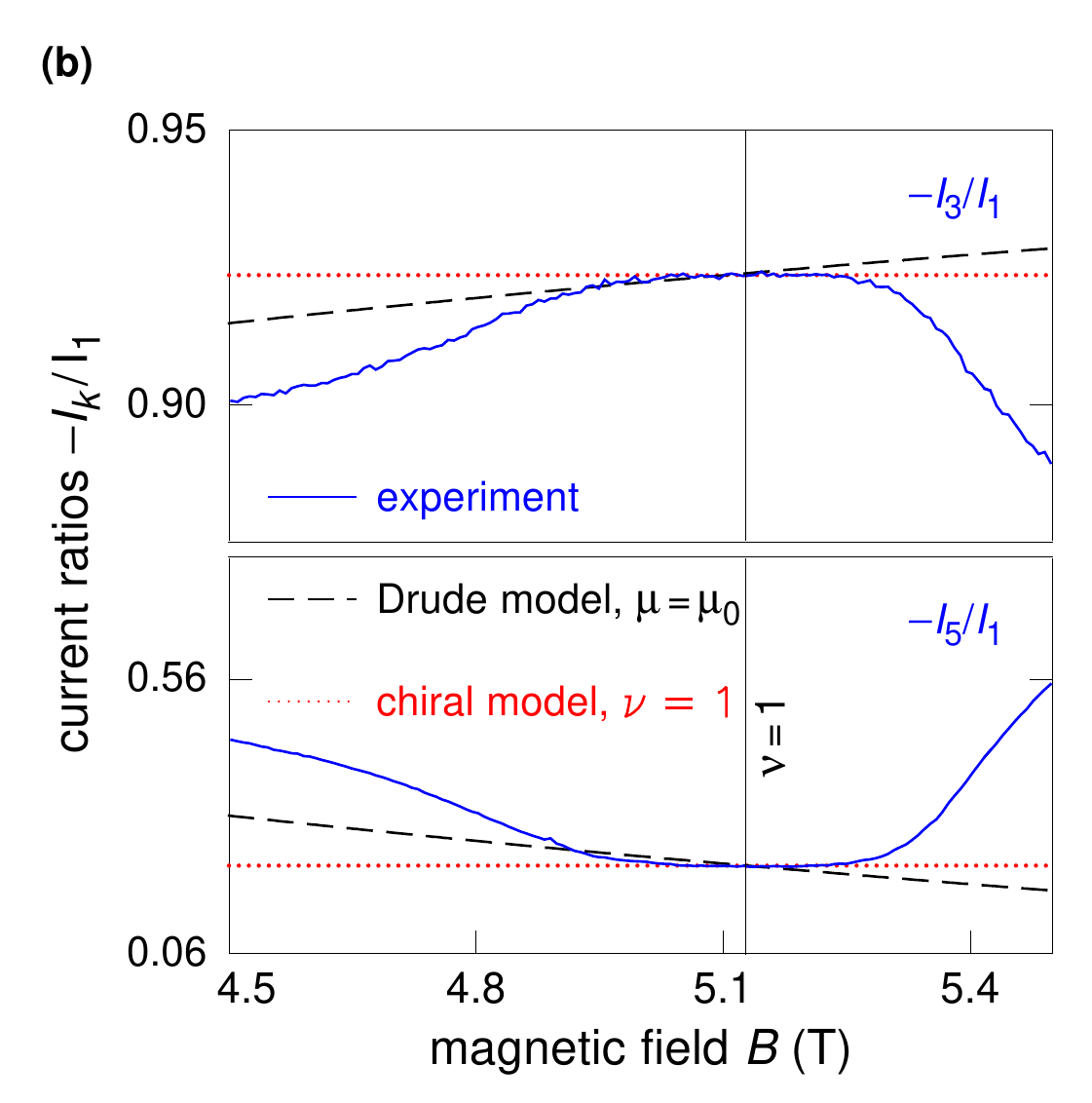}
\caption{Three-terminal measurement with symmetric circuit, cf.\ inset in (a). A voltage of $V=1\,$mV is applied to contact 1, while contacts 3 and 5 are connected to ground via current amplifiers; contacts 2, 4, and 6 are electrically floating. $\vec B$ points from the surface into the 2DES plain such that $I_{xy}$ is bent towards contacts 2 and 3.
(a) Current ratios $-I_3/I_1$ and $-I_5/I_1$ as a function of magnetic field $B$ (solid blue lines); Drude model prediction of \eq{currents_Drude_nterminal} assuming constant $\mu=\mu_0$ (dashed black line); fit of Drude model prediction of \eq{currents_Drude_nterminal} with $\mu(B)$ as free parameter (dotted red lines).
(b) enlargements near $\nu=1$ showing experimental data, the Drude model prediction for $\mu=\mu_0$ and the model prediction of \eq{currents_coherent} for coherent and chiral current at $\nu=1$ (dotted red lines).
}
\label{fig:3T_measurements}
\end{figure*}

We start our quantitative comparison between measurements and models for the three-terminal measurements already presented in \fig{fig:overview}{} above. Here, the grounded contacts 3 and 5 are arranged in a mirror symmetric geometry in respect to the horizontal axis through contact 1. In \fig{fig:3T_measurements}{a},
\begin{figure}[htb]
\includegraphics[width=0.8\columnwidth]{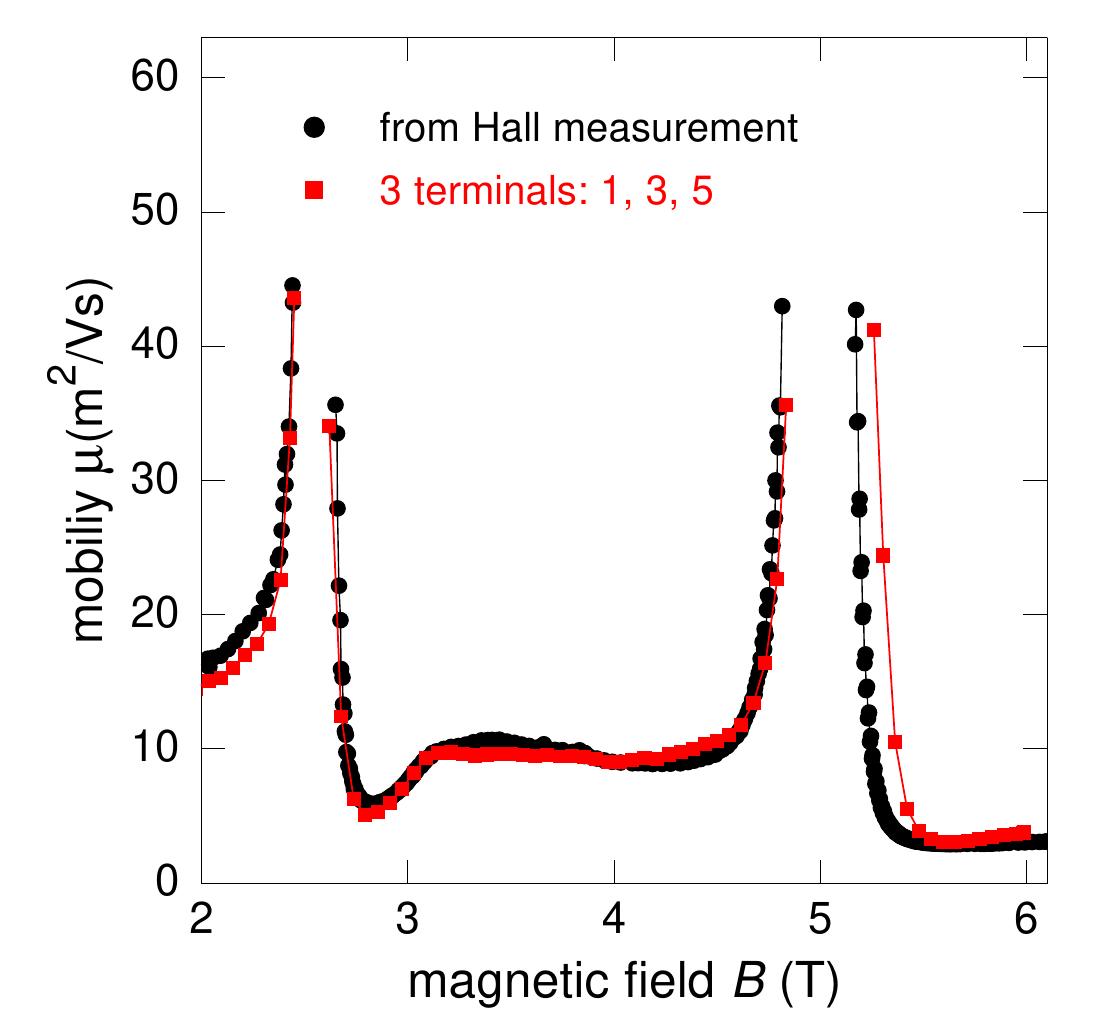}
\caption{Mobility $\mu(B)$ as determined from SdH-oscillations of the longitudinal resistance between contacts 2 and 3 (black dots) in comparison with $\mu(B)$ found by fitting \eq{currents_Drude_nterminal} to the three-terminal measurements shown in Fig.\ \ref{fig:3T_measurements} (red squares).}
\label{fig:mobilities}
\end{figure}
\begin{figure*}[tb]
\includegraphics[width=0.33\textwidth]{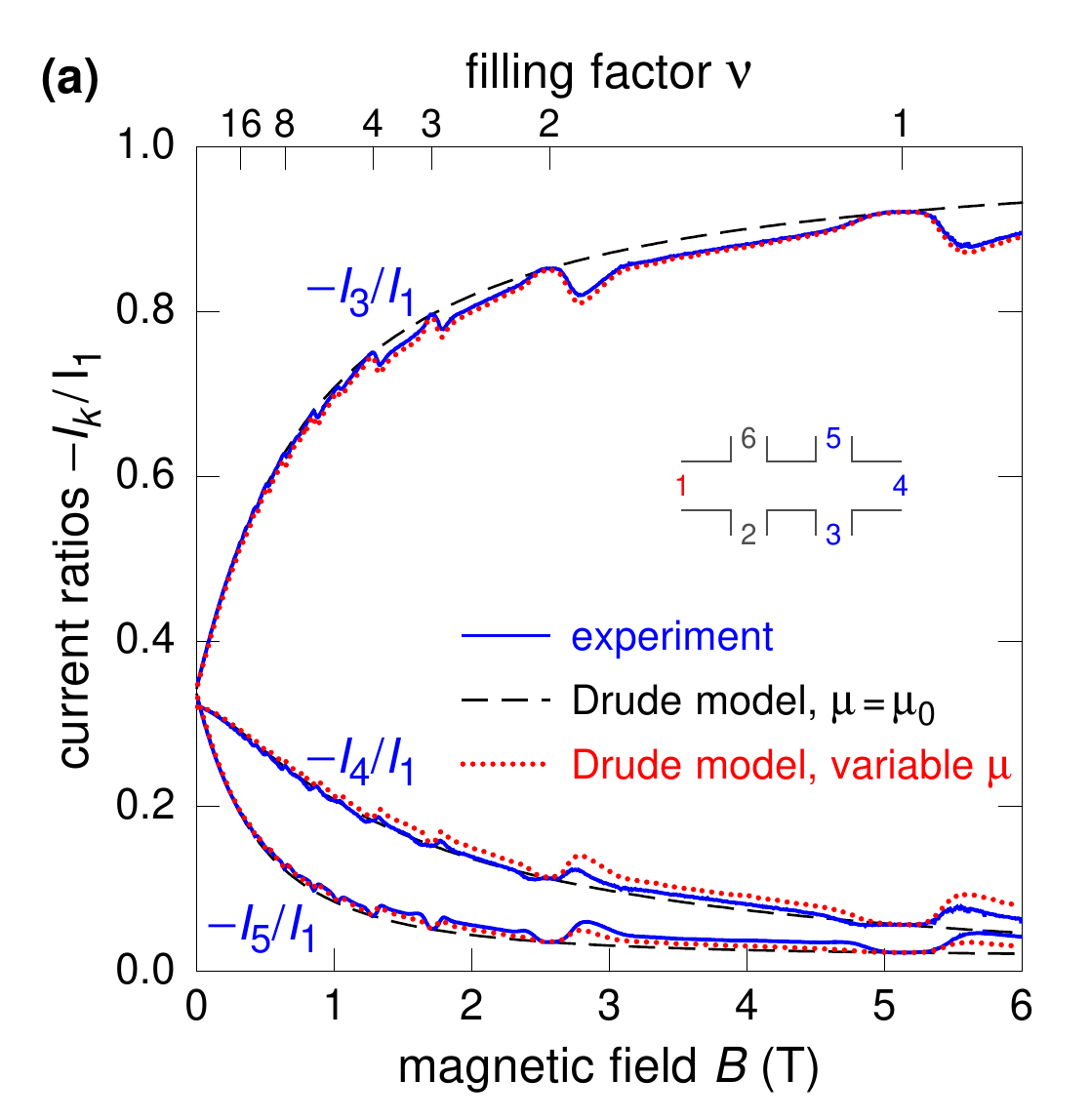}
\includegraphics[width=0.33\textwidth]{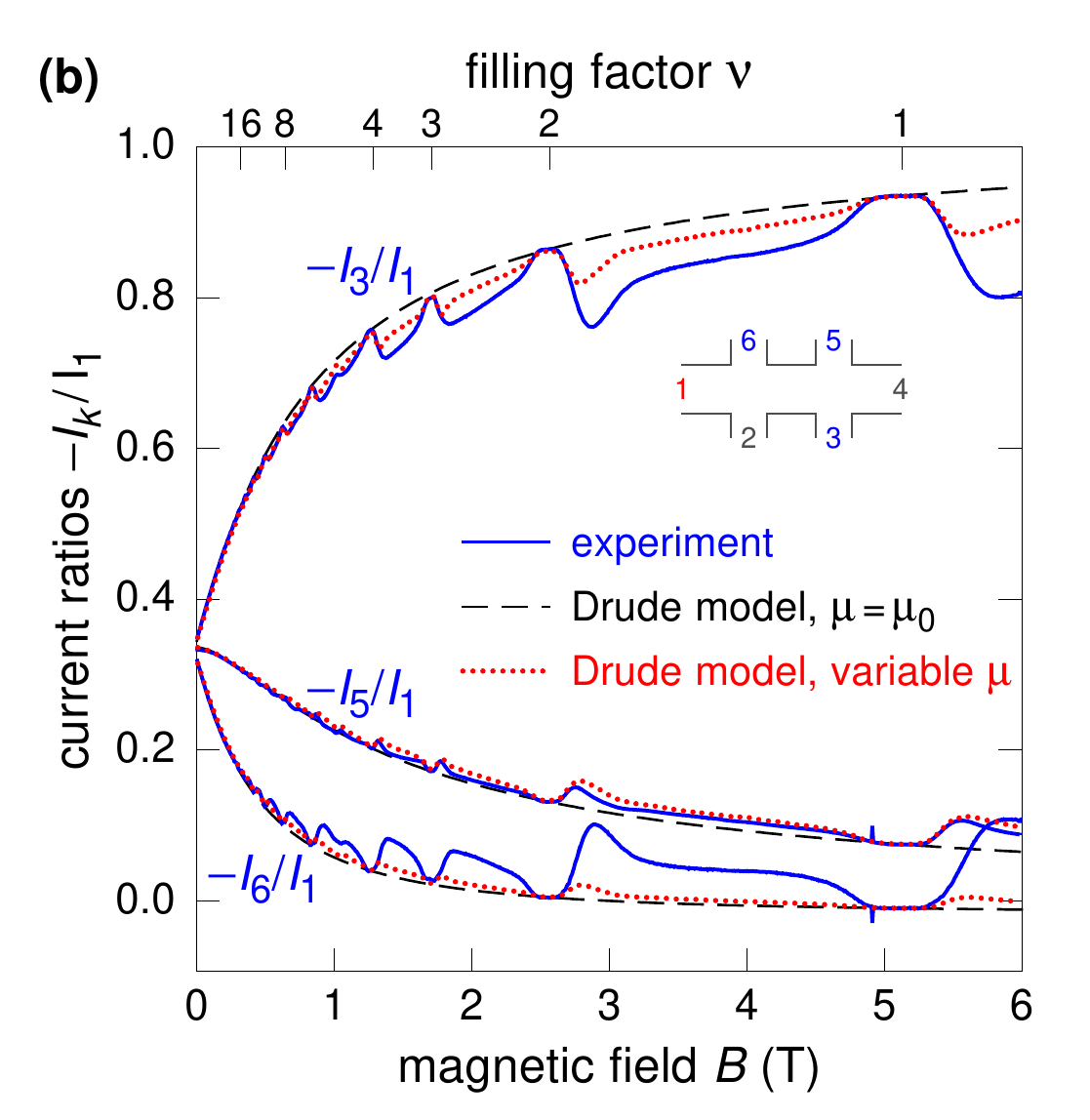}
\includegraphics[width=0.33\textwidth]{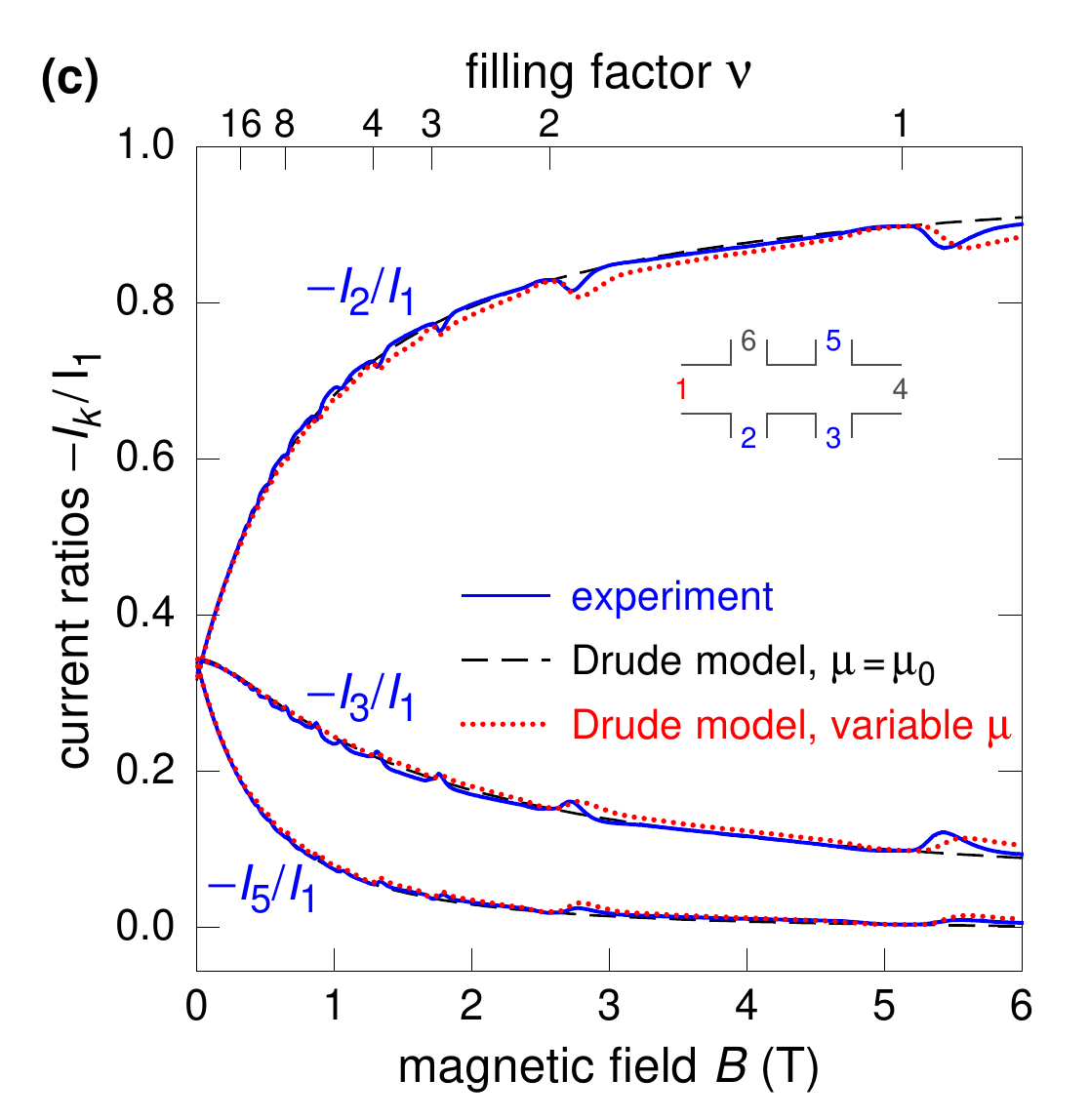}
\caption{Four-terminal measurements for three different contact configurations; the blue numbers in the insets indicate grounded contacts, black numbers indicate floating contacts and the external voltage is applied to contact 1 (red); further details as in Fig.\ \ref{fig:3T_measurements};
(a) symmetric four-terminal circuit;
(b) and (c) asymmetric configurations.}
\label{fig:4T_measurements}
\end{figure*}
we present as solid (blue) lines the measured current ratios $-I_3(B)/I_1(B)$ and $-I_5(B)/I_1(B)$. As already discussed above, $-I_3/I_1$ increases with $B$ while $-I_5/I_1$ decreases accordingly, such that $-I_3(B)/I_1(B)-I_5(B)/I_1(B)=1$. The dashed black line indicates the prediction of the Drude model according to \eq{currents_Drude_nterminal} for $n=3$, if we use the constant mobility $\mu(B)=\mu_0=395\,\textrm{m}^2/\textrm{Vs}$ that we measured at $B=0$.

In the classical limit for $B\lesssim 0.5$\,T the Drude model agrees well with our measurement. As we increase $B$ further, the Drude model still predicts the correct trend but as long as we use a constant $\mu(B)=\mu_0$ (black dashed lines), we observe growing deviations. These are clearly visible in the enlargement around the first plateau at $\nu=1$ in \fig{fig:3T_measurements}{b}. The red dotted lines in \fig{fig:3T_measurements}{b} are the prediction of the chiral model according to \eq{currents_coherent} [corresponding to \eq{currents_Drude_nterminal} for $\mb=\infty$] and a constant filling factor $\nu=1$. It perfectly fits the values of the current ratios along the $\nu=1$ plateau of the quantized Hall resistance. An exact match between measurements and model beyond the plateaus can be achieved by using $\mu(B)$ as a free fit parameter in \eq{currents_Drude_nterminal}. For the dotted red line in \fig{fig:3T_measurements}{a}, we used the magnetic field dependent values of $\mu(B)$, which are plotted as red squares in \fig{fig:mobilities}{}.

For comparison, we have also determined $\mu(B)$ from the SdH oscillations of a standard longitudinal resistance measurement, see the black dots in \fig{fig:mobilities}{} \footnote{The slight deviations between the two are likely related with local changes in the sample during thermal cycles of the He-3 evaporation cryostat to $T>4\,$K between the two measurements.}. The general agreement between the $\mu(B)$ values obtained from two completely independent measurements, namely longitudinal resistance measurements versus fitting \eq{currents_Drude_nterminal} to our three-terminal current measurements, strongly supports the validity of our model.

The two regions of divergent $\mu(B)$ in \fig{fig:mobilities}{} correspond to the plateau regions with the local filling factors $\nu=2$ and $\nu=1$. Here, $I_{xx}$ vanishes supporting $\mb\to\infty$. Away from the plateau regions, in \fig{fig:mobilities}{} we find mobility values strongly reduced even compared to $\mu_0$. This indicates a transition from ballistic transport on the plateaus to diffusive transport between the plateaus. The difference between the Drude model predictions for $\mu=\mu_0$ [dashed line in \fig{fig:3T_measurements}{a}] and the measured currents [blue line] is precisely the ohmic current contribution $I_{xx}$, which becomes clearly detectable for $\mb<20$.

To understand the meaning of the perfect agreement of the Drude model fits based on a variable $\mu(B)$ it is useful to mention, that for a large enough mobility, here $\mb\gtrsim20$, the Ohmic term (first term) of \eq{currents_Drude_nterminal} becomes too small to be detected. Neglecting $I_{xx}$ in \eq{currents_Drude_nterminal} and using $(\nu G_0)^{-1}=\rh(B)$, then makes \eq{currents_coherent} [chiral model] and \eq{currents_Drude_nterminal} [Drude model] mathematically identical.

So-far we have discussed  three-terminal current measurements. With the knowledge of the contact resistances of our sample, we can quantitatively compare the experiments with model predictions. Between the plateaus, the mobility is strongly reduced and the Drude model fits perfectly to our data, indicating non-chiral diffusive transport. Regarding the quantized plateau regions, our three-terminal measurements are clearly consistent with \eq{currents_coherent}, describing chiral current flow, if we assume a constant integer filling factor (for the entire plateau). It confirms, that for the quantized Hall plateaus the current flows within regions of constant integer filling factor, while the bulk filling factor (averaged over the width of the Hall bar) continuously decreases proportional to $1/B$. Importantly, any additional non-zero transmission coefficients in \eq{Gij} due to non-chiral currents would modify the prediction of \eq{currents_coherent}. Clearly, the fact that \eq{currents_coherent} perfectly fits our data for the plateaus with $\mb\to\infty$ (without the need of additional non-zero transmission coefficients) suggests chiral transport for the quantized Hall plateaus.

\subsection{Four-terminal measurements}\label{sec:4T-measurements}

Three-terminal measurements are special, because if a model curve agrees with one current branch, say $-I_3/I_1$ in \fig{fig:3T_measurements}{}, because of $I_1+I_3+I_5=0$ it will also agree with the second branch, here $-I_5/I_1$. Clearly, this is not the case if we add additional terminals. Then, any inaccuracies of our model will be visible as deviations between measurements and model predictions. In that sense, experiments with four or more terminals provide a stricter test of our models compared to the three-terminal measurements above.

In \fig{fig:4T_measurements}{}, we present current ratios $-I_k/I_1$ for three different four-terminal measurements (blue solid lines). Again, we measure the currents into contacts $k=1,3$ and 5, but now we ground an additional contact, namely contact 4 in \fig{fig:4T_measurements}{a}, contact 6 in (b), and contact 2 in (c). If we assume $\mu(B)=\mu_0$,  \eq{currents_Drude_nterminal} reproduces the overall dependences $I_k(B)$ with a similar accuracy and similar deviations as it did for the three-terminal measurement (dashed black lines).

The dotted red lines correspond to model curves of \eq{currents_Drude_nterminal} using the $\mu(B)$ plotted in \fig{fig:mobilities}{} and determined by fitting the three-terminal measurement shown in \fig{fig:3T_measurements}{}. As it was the case for the three terminal measurements, for the plateau regions, corresponding to diverging $\mu(B)$, the agreement between theory and the measured currents is perfect.  This observation supports our previous conclusion, that in the plateau regions the current flow is chiral, while away from the plateaus, the chiral model fails.

Noticeable, away from the plateaus, where the current flow is diffusive, the agreement between theory curves and measured data is no longer perfect. Thereby, that the best agreement occurs for the symmetric contact configuration in \fig{fig:4T_measurements}{a}. This behavior suggests, that our model is inaccurate in the diffusive transport regime. Thereby, the deviations depend on the details of the contact configuration.

In conclusion, our four-terminal measurements confirm the main findings of our three-terminal measurements: (1) Away from the plateaus, the transport is not chiral. (2) For the plateau regions, the agreement between our four-terminal measurements and the chiral and coherent model supports chiral transport.

\subsection{Limits of our model in the diffusive regime}\label{sec:limits}

The deviations between the prediction of \eq{currents_Drude_nterminal} and our measurements in the diffusive regime point to an incomplete model for $I_{xx}$. The problem is not related with the different contact resistances and offset voltages of the current amplifiers, because they are taken into account in \eq{currents_Drude_nterminal}. Instead, it is related with the mesoscopic dynamics of the electrons inside the Hall bar. In \sect{sec:Drude-nterminal} we determined the distribution of $I_{xx}$ into various contacts based on the contact resistances and the Hall resistance in series, while treating the Hall bar itself as a current node. To do this, we assumed a homogeneous current density distribution across the Hall bar. However, in particular near the intersections between the Hall bar and its contacts the distributions of the electric field and the current density become locally inhomogeneous. We conjecture, that such local inhomogeneities near the intersections cause the observed deviations between measured data and model curves in \fig{fig:4T_measurements}{}. Note that the inhomogeneities affect the distribution of $I_{xx}$ into various contacts, while  $I_{xy}$ is nevertheless bent into the contacts due to the charge accumulated along the edge of the Hall bar. A more realistic model would require a suitable numerical calculation, e.g.,  based on a path integral formalism with realistic boundary conditions.

\subsection{Experimental confirmation of non-chiral currents between plateaus}

\begin{figure}[th]
\includegraphics[width=0.8\columnwidth]{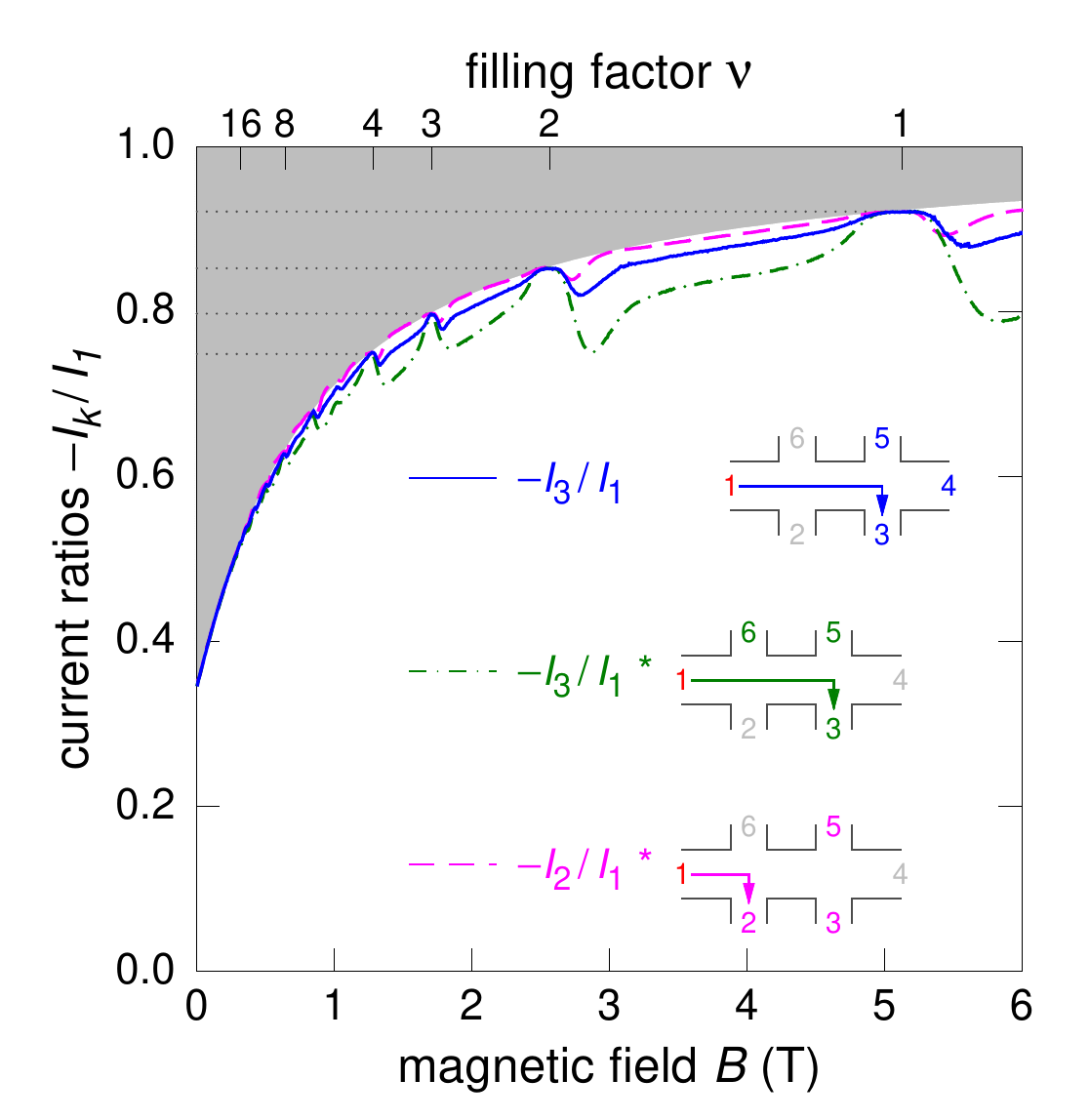}
\caption{Comparison of the respective largest measured current contributions flowing into a grounded contact for three different configurations of four-terminal measurements. The dotted horizontal lines depict the predicted current ratios for filling factors $1\le\nu\le4$. The current ratios $I_k\,/\,I_1$ marked with a $*$ are slightly scaled to account for the slightly different contact resistances, such that all three theory curves according to \eq{currents_Drude_nterminal} with $\mu=\mu_0$ are identical and follow the edge between shaded and white backgrounds.
Sketches: Arrows indicate the displayed current branches; colored numbers indicate current carrying and gray numbers floated contacts.}
\label{fig:currents_vgl}
\end{figure}

In \fig{fig:currents_vgl}{},
we plot for each of the three four-terminal measurements shown in \fig{fig:4T_measurements}{} only the branch $I_k/I_1$ corresponding to the largest current. In a chiral model, this is the current flowing into the first grounded contact after emission from contact 1.  
We slightly scaled the data (based on the values provided in Table \ref{table:parameters}) to compensate for small variations in the contact resistances. After scaling, the three curves are identical in the classical regime ($B<0.5\,$T) as well as on the plateaus of the QHE. The line linking shaded and white backgrounds corresponds to the prediction of \eq{currents_Drude_nterminal} for constant $\mu=\mu_0$ [equal to the dashed black lines in \fig{fig:4T_measurements}{a}].

Compared to the model prediction, between the plateaus the current is reduced for all three geometries, which above we already interpreted in terms of a finite contribution of the non-chiral $I_{xx}$. Because the various contact resistances of our sample are almost identical and the number of grounded contacts is identical, for a purely chiral current flow the  branches of the largest current $I_k(B)/I_1$ should be independent of the details of the contacts configuration. While this is the case on the quantized Hall plateaus, away from the plateuas $I_k(B)/I_1$ varies substantially between configurations. This can only be understood, if we allow for non-chiral currents between the quantized Hall plateaus.

This conclusion experts additional support by a closer look at the amounts of current reduction, based on its systematic dependence on the relative distances between contact 1 and the grounded contacts on both sample edges in a way that contradicts chiral current flow. Namely, the current supression into the contact with the largest current is the greater the closer another grounded contact is to the biased contact 1.

\subsection{Formulation of the proof of chiral current flow for quantized Hall resistance}\label{sec:chirality_proof}

So far, we have proven non-chiral current flow between the quantized plateaus, while for the quantized plateaus we have merely stated that our experimental results are consistent with chiral current flow. However, our analysis allows a stronger statement, namely that the agreement between the chiral model and our measurements is a prove of chiral current flow, if we accept that \eq{Gij} describes chiral current flow.

The proof is based on the coherent and chiral model introduced in \sect{sec:coherent_model}. Based on the Landauer-Büttiker formalism it describes the current division into $n$ contacts in terms of the mutual conductance coefficients $G_{kj}$ between the contacts. In this model, chiral current flow corresponds to the specific values of conductance coefficients given by \eq{Gij}. A non-chiral behavior would correspond to different values, including additional non-zero conductance coefficients in \eq{Gij}. These would inevitably result in different plateau values of the predicted current ratios. 

In all our measurements including three or four terminals shown in \fig{fig:3T_measurements}{}, \fig{fig:4T_measurements}{} and \fig{fig:currents_vgl}{}, we instead see a perfect agreement between our chiral model prediction and the measured plateau values of the current ratios. This is, in particular, evident when comparing the horizontal dotted lines in \fig{fig:3T_measurements}{b} and \fig{fig:currents_vgl}{}, which indicate the predictions of \eq{currents_coherent} for the corresponding integer filling factors, with the corresponding measured plateau values.

\section{Summary and Outlook}\label{sec:summary}

We have performed multiterminal current measurements of a Hall bar in the regime of the QHE. We find that the current flow is chiral in the magnetic field regions of the quantized Hall resistance plateaus but becomes non-chiral in between the plateaus. Our result is based on the direct comparison of the measured current distribution into various contacts (three or four terminals) with two limiting models, each providing a unique physical meaning. The classical Drude model serves us as the basis to describe non-chiral diffusive transport, while our second model is based on the generic chiral and coherent transmission problem. The comparison of different contact configurations for more than three terminals allows for a direct experimental proof of non-chiral currents, which confirm our model based results. Our work contributes to the understanding of the current flow inside mesoscopic devices in the regime of the QHE. It thereby confirms the assumption of chiral current flow frequently used for the case of quantized plateaus of the Hall resistance.

\section*{Acknowledgement}

The authors thank Piet Brouwer for help with the coherent model and Afif Siddiki and Lutz Schrottke for fruitful discussions. This work was funded by the Deutsche Forschungsgemeinschaft (DFG, German Research Foundation) -- 218453298.

\section*{Contributions of the authors}

V.\,Y.\,U.\ supplied the wafer. S.\,S.\ and M.\,K.\ performed the measurements. S.\,S.\ and S.\,L.\ analyzed the data. S.\,L.\ derived the models. S.\,S.\ and S.\,L.\ wrote the article.

\section*{References}

%

\appendix
\section{Currents predicted by our coherent and chiral model}\label{app:a}

To derive \eq{currents_coherent} in \sect{sec:coherent_model} describing chiral and coherent current within the inner circle of \fig{fig:nT-sketch}{}, rewrite \eq{LBF},
\begin{equation}\nonumber
I_k=\sum_{j=1}^{n} G_{kj}V_j'=(V_k'-V_k)/R_k\,.
\end{equation}
Its left hand side describes the coherent current flowing inside the inner circle of \fig{fig:nT-sketch}{} according to the Landauer-Büttiker formalism. Its right hand side describes the interfaces between the coherent inner circle with the non-coherent world by relating the same currents with the contact resistances based on Ohm's law. Requiring unitarity of the coherent scattering matrix demands $\sum_k G_{kj}=\sum_j G_{kj}=0$. Additionally requiring chiral current flow, such that current flows only from contact $1\to2$, $2\to3$, \dots, $n-1\to n$, $n\to1$ yields our conductance matrix [equivalent to \eq{Gij}]
\begin{equation}
\hat G = \nu G_0
\begin{pmatrix}
1 & 0 & \cdots & 0 &-1 \\
-1& 1 & 0 & \cdots & 0 \\
0 &-1 & 1 & 0 & \cdots \\
\vdots & \ddots & \ddots & \ddots & \vdots \\
0 & \cdots & 0 & -1 & 1 \\
\end{pmatrix}\,,\nonumber
\end{equation}
where we used $\nu G_0$ for the conductance between pairs of contacts. Within the Landauer-Büttiker formalism, $\nu$ stands for the number of coherent channels, each contributing with the one-dimensional conductance quantum, $G_0=e^2/h$, to the current. For the case of the QHE, $\nu$ corresponds to the filling factor and the Hall resistance is $\rh=1/\nu G_0$. In our experiments, the voltages, $V_k$, are applied to the contacts, while we measure the currents, $I_k$, flowing into each contact. In order to predict these currents, we first determine the voltages, $V'_k$, at the nodes between the coherent inner circle of \fig{fig:nT-sketch}{} and the ohmic contacts. We can rewrite the right hand side of \eq{Gij} as
\begin{equation}\label{LBF2}
\sum_{j=1}^n g_{kj} V'_j = -\frac{V_k}{\nu G_0 R_k}\equiv -w_k\,,
\end{equation}
where we introduced the dimensionless conductance matrix
\begin{equation}
\hat g =
\begin{pmatrix}
-\gamma_1 & 0 & \cdots & 0 & 1 \\
1 & -\gamma_2 & 0 & \cdots & 0 \\
0 & 1 & -\gamma_3 & 0 & \cdots \\
\vdots & \ddots & \ddots & \ddots & \vdots \\
0 & \cdots & 0 & 1 & -\gamma_n \\
\end{pmatrix}\nonumber
\end{equation}
whith $\gamma_k=1+(\nu G_0 R_k)^{-1}$. Sorting  \eq{LBF2} for $V'_k$ results in a unique solution of the linear equation system
\begin{eqnarray}
V'_1 &= \frac 1 {\gamma_1} \left(V'_n+w_1 \right) &\nonumber\\
V'_2 &= \frac 1 {\gamma_2} \left(V'_1+w_2 \right) &= \left (V'_n + P_0 w_1 + P_1 w_2 \right) / P_2 \nonumber\\
V'_3 &= \frac 1 {\gamma_3} \left(V'_2+w_3 \right) &= \left (V'_n + P_0 w_1 + P_1 w_2 + P_2 w_3\right) / P_3 \nonumber\\
\dots & \dots & \nonumber\\
V'_n &= \frac 1 {\gamma_n} \left(V'_{n-1}+w_n \right) &=  \sum_{j=1}^nP_{j-1}w_j / \left(P_n-1\right)\nonumber    \,,
\end{eqnarray}
where we introduced the product
\begin{equation}
P_k=\left\{ \begin{matrix} \prod_{j=1}^k \gamma_j & \text{for} & j=1,2,3,\dots \\ 1 & \text{for} & j=0\end{matrix}\right.\,.\nonumber
\end{equation}
With the additional abbreviation
\begin{equation}
S_k=\sum_{j=1}^k P_{j-1}w_j\nonumber
\end{equation}
we find
\begin{equation}
V'_k=\frac 1 {P_k}\left(\frac{S_n}{P_n-1} + S_k \right)\,.\nonumber
\end{equation}
In the last step, we use $I_k=\left(V'_k-V_k\right)/R_k$ to express the current into the $k$th contact in terms of the contact resistances $R_k$ and the applied voltages $V_k$:
\begin{equation}
I_k=\frac{1}{R_k}\left[\frac{1}{P_k}\left(\frac{S_n}{P_n-1}+S_k\right)-V_k \right]\,,\nonumber
\end{equation}
which is \eq{currents_coherent}.

\section{Diffusive currents predicted by the Drude model}\label{app:b}

\begin{figure}[ht]
\includegraphics[width=.8\columnwidth]{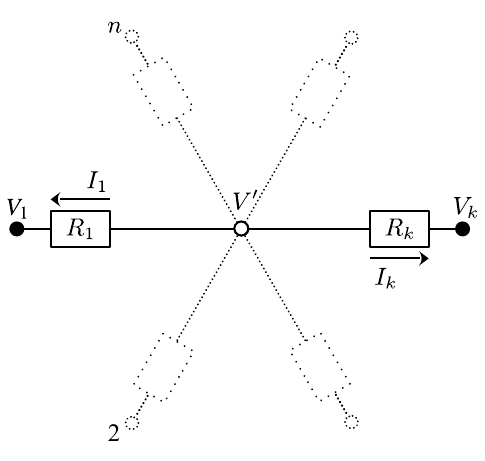}
\caption{Simplified circuit diagram of an $n$-terminal device at $B=0$ neglecting the resistance $R_0$ of the 2DES. It describes the current contribution $I_{xx}$ of the Drude model.}
\label{fig:nT-sketch_Ohm}
\end{figure}

Integration over the sample width of the current density in \eq{jx} results in
\begin{eqnarray}
I_k&=&I_{xx}^{(k)}+I_{xy}^{(k)}\\
   &\equiv&\frac{1}{1+\mbq}I_x^{(k)}+\frac{\mbq}{1+\mbq}I_y^{(k)}
\end{eqnarray}
For determining the ohmic currents $I_x^{(k)}$, we assume $V'_1=V'_2=\dots=V'_n\equiv V'$ in \fig{fig:nT-sketch}{}, which corresponds to the neglect of the resistance $R_0$ of the 2DES. Our simplified circuit diagram takes the form of that in cf.\ \fig{fig:nT-sketch_Ohm}{}.
This simplifying ohmic circuit is described by the $n$ equations
\begin{equation}
I_k=\left(V'-V_k\right)/R_k\text{ for }k=1,2,\dots,n\nonumber
\end{equation}
and Kirchhoffs current law $\sum_{j=1}^n I_j=0$. Combining both, we find
\begin{equation}
V'=\frac{\sum_{j=1}^n V_j/R_j}{\sum_{j=1}^n 1/R_j}\nonumber
\end{equation}
and
\begin{equation}
I_x^{(k)}=\frac{1}{R_k}\left(\frac{\sum_{j=1}^n V_j/R_j}{\sum_{j=1}^n 1/R_j}-V_k\right)\,,\nonumber
\end{equation}
which explains the $I_{xx}$ in \eq{currents_Drude_nterminal}.
For $B\simeq0$ we measured $R_0\simeq50\,\Omega$, justifying its neglect compared to the ohmic resistances of approximately 2.5\,k$\Omega$ per contact. However, with increasing magnetic field the mobility decreases and strongly oscillates (as a function of $B$), such that $R_0(B)$ becomes similar to the contact resistances at its minima. Then, the detailed current density distribution matters for the partitioning of $I_{xx}^{(k)}$ into the various contacts, in particular, if the path ways into various contacts correspond to different effective resistances of the Hall bar. To account for this contribution of the Hall bar itself to the resistance in the diffusive regime, we increase the contact resistances accordingly. To achieve this, we defined a geometry parameter $a_k$ and use $R_k\to R_k+a_k/[\ns e \mu(B)]$ in the model calculations of the ohmic contributions, see Table \ref{table:parameters}. The correction is negligible for $B=0$, but increases for growing $B$. The parameter $a_k$ is a fairly rough estimate to account for geometry details and is not sufficient to correct for the asymmetries of the four-terminal configurations in \fig{fig:4T_measurements}{}. This could explain, why between the plateaus our four-terminal measurements presented in \fig{fig:4T_measurements}{} show a relative poor quantitative agreement with our model curves. A more accurate model would require numerical path integral calculations.

\begin{figure}[ht]
\includegraphics[width=.8\columnwidth]{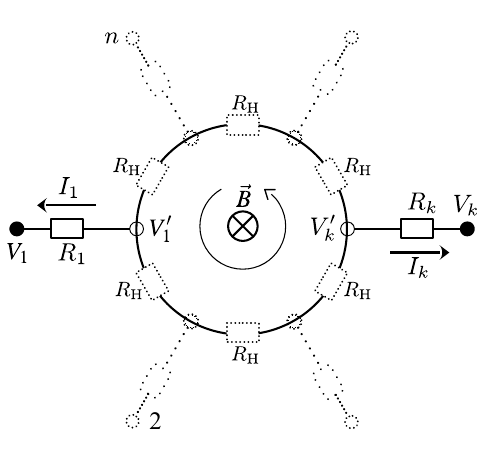}
\caption{Simplified circuit diagram of an $n$-terminal device including the Hall resistance to describe the current contribution $I_{xy}$ of the Drude model. The inner circle indicates directional but not chiral current flow through the 2DES from contact to contact. Each contact $k=1,...,n$ has an ohmic resistance $R_k$ and a voltage $V_k$ applied in respect to the electrical ground.  $V_k'$ are the voltages resulting at the intersections between the ohmic contacts and the coherent inner circle. The magnetic field vector pointing into the plane of the 2DES gives rise to a right handed chirality of a free charge which translates into a left handed chirality of the device, cf.\ circled arrow. }
\label{fig:nT-sketch_app}
\end{figure}

The currents $I_y^{(k)}$ are bent from contact to contact by the charged sample edges as indicated in \fig{fig:nT-sketch_app}{}. Thereby, according to \eq{jx}, inside of the 2DES between contacts, the current $I_{xy}$ is additionally impeded by the Hall resistance \rh. According to the Drude model, the current contribution $I_{xy}$ flows unidirectional and, thus, is not chiral. However, it flows only between adjacent contacts in a pre-defined direction.
The model describing $I_{xy}^{(k)}$ is mathematically equivalent with our coherent and chiral model, if we use $\nu G_0=1/\rh$. However, while the coherent model applies for integer filling factors only, the pre-conditions of the Drude model exclude the limit of integer $\nu$.

\section{Hall measurements}\label{app:c}

\begin{figure}[htb]
\includegraphics[width=1\columnwidth]{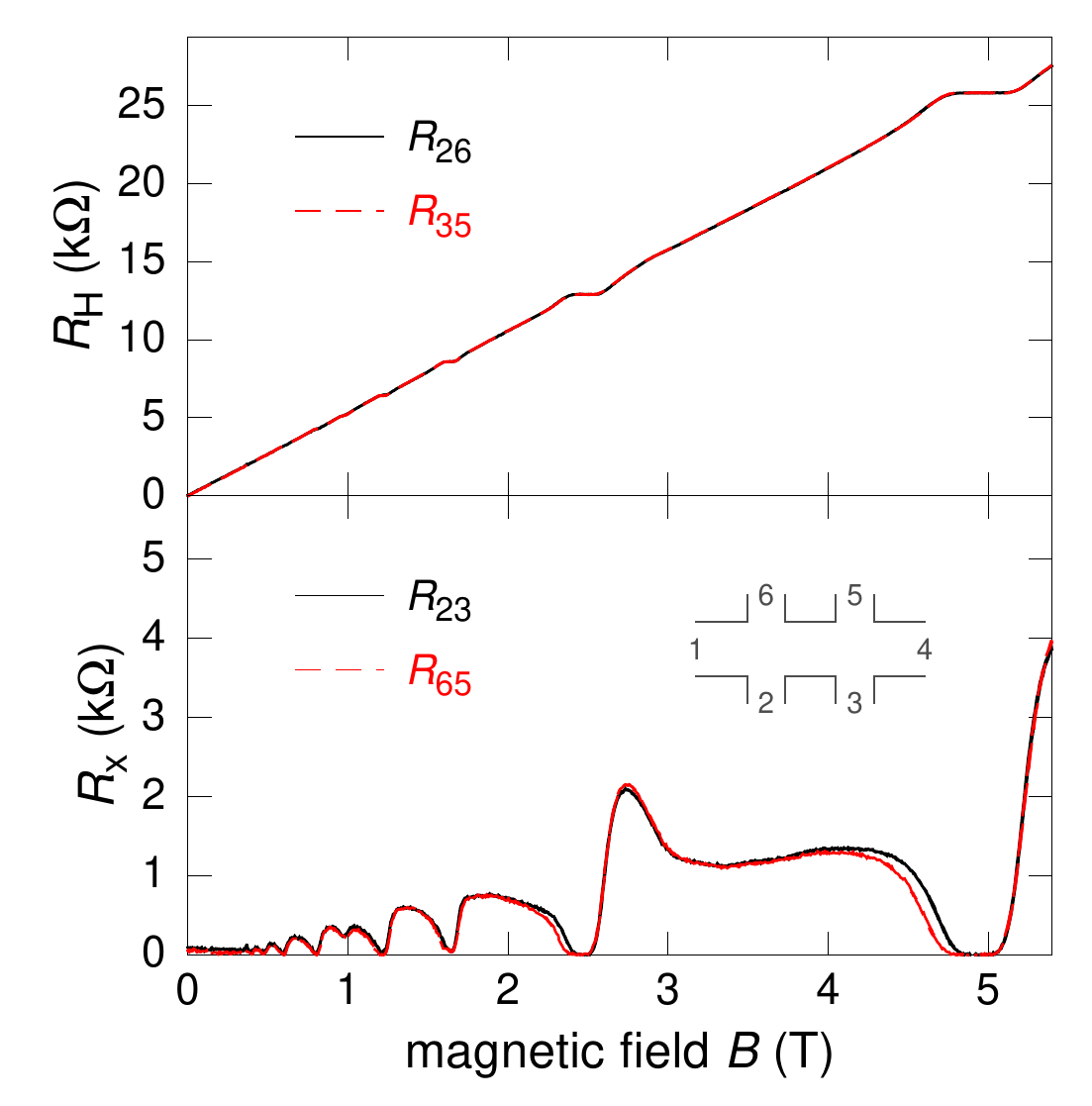}
\caption{Standard Hall measurements, $R_{ij}=V_{ij}/I_{41}$ of our Hall bar at the temperature of $T\simeq300\,$mK, current applied between contacts 1 and 4, cf.\ inset. Hall resistances, $R_\text{H}(B)$, in the top panel and longitudinal resistances, $R_\text{x}(B)$, in the bottom panel.}
\label{fig:Hall_measurements}
\end{figure}

In \fig{fig:Hall_measurements}{},
we present standard four-terminal measurements, namely the Hall resistances, $R_\text{H}$, and longitudinal resistances, $R_{x}$, as a function of $B$ while a constant current of 100\,nA was imposed between contacts 4 and 1, cf.\ inset. We determined the mobility shown as black dots in \fig{fig:mobilities}{} from these data using $\mu=\frac{l}{b}\frac{R_\text{H}}{R_x}\,/B$, where $l\simeq\,20\,\mu$m and $b\simeq10\,\mu$m are the distance between contacts 2 and 3 (6 and 5) and the width of the Hall bar. The Hall resistances $R_{26}$ and $R_{35}$ as well as the longitudinal resistances $R_{23}$ and $R_{65}$ are almost identical, respectively, although they are measured between different pairs of contacts. This result suggests, that the carrier density in our Hall bar is fairly symmetric in regard to the four contacts, i.e., homogeneous away from the edges. It supports our assumption of a homogeneous carrier density in the main article. A well accepted method to determine the degree of homogeneity inside a Hall bar are Dingle plots of the SdH oscillations in the low magnetic field regime \cite{Coleridge1991}. However, due to the relatively small size and high quality of our Hall bar, all of our SdH oscillations in \fig{fig:Hall_measurements}{} are asymmetric with minima at $R_x=0$ and show spin splitting even for $B<1\,$T, making an analysis with a Dingle plot according to Ref.\ \cite{Coleridge1991} unpractical.

The slight deviations between the longitudinal resistances, $R_{23}$ and $R_{65}$, in \fig{fig:Hall_measurements}{} are not reproducable and  might be related with the fact that we measured the voltages at the four contacts with four separate volt meters. A slight shift in the offset of one of them between measurements might cause the observed differences.

\end{document}